# Context-Aware Misbehavior Detection Scheme for Vehicular Ad Hoc Networks using Sequential Analysis of the Temporal and Spatial Correlation of the Cooperative Awareness Messages

Fuad A. Ghaleb[ab], Mohd Aizaini Maarof[a*], Anazida Zainal[a], Murad Rassam[c], Faisal Saeed[d], Mohammed Alsaedi[d]

[a]Cybersecurity X Lap, Information Assurance and Security Research Group, School of Computing, Faculty of Engineering, Universiti Teknologi Malaysia, 81310, Johor, Malaysia
[b]Department of Computer and Electronic Engineering, Sana'a Community College, Sana'a 5695, Yemen
[c]Faculty of Engineering and Information Technology, Taiz University, Taiz 6803, Yemen
[d]College of Computer Science and Engineering, University of Taibah, Medina, Saudi Arabia, Saudi Arabia

*Email:* aizaini@utm.my, aagfuad2@live.utm.my, fuadeng@gmail.com

Abstract

Vehicular ad hoc Networks (VANETs) are emerged mainly to improve road safety, traffic efficiency, and passenger comfort. The performance of most VANET applications and services relies on the availability of accurate and recent mobility-information, so-called Cooperative Awareness Messages (CAM), shared by neighboring vehicles. However, misbehaving in terms of sharing false mobility information can disrupt any potential VANET application. Because cryptographic solutions in VANET are expensive, complicated, and vulnerable to internal misbehavior, security lapses are inevitable. Therefore, misbehavior detection is an important security component. Unfortunately, existing misbehavior detection solutions lack considering the high dynamicity of vehicular context which leads to low detection accuracy and high false alarms. The use of predefined and static security thresholds are the main drawbacks of the existing solutions. In this paper, a context-aware misbehavior detection scheme (CAMDS) is proposed using sequential analysis of temporal and spatial correlation of neighboring vehicles' mobility information. A dynamic context reference is constructed online and timely updated using statistical techniques. Firstly, the Kalman filter algorithm is used to track the mobility information received from neighboring vehicles. Then, the innovation errors of the Kalman filter are utilized to construct a temporal consistency assessment model for each neighboring vehicle using Box-plot. Then, the Hampel filter is used to construct a spatial consistency assessment model that represents the current context reference model. Similarly, plausibility assessment reference models are built online and timely updated using the Hampel filter and by utilizing the consistency assessment reference model of neighboring information. Finally, a message is classified as suspicious if its consistency and plausibility scores deviate much from the context reference model. The proposed context-aware scheme achieved a 73% reduction in false Alarm rate while it achieves a 37% improvement in the detection rate. This proves the effectiveness of the proposed scheme compared with the existing static solutions.

**Keywords**, Context-Aware Misbehavior Detection, vehicular ad hoc networks (VANETs), vehicular networks, Cooperative Intelligent Transportation System (cITS), Cooperative Awareness Messages.

1. Introduction

According to the World Health Organization (WHO), road collisions are increasing, and they are being expected to be the fifth leading causes of death by 2030 [1]. Annually, 1.25 million people lose their lives on roads with forty times more suffering injuries. Accidents also are the leading causes of traffic congestion, which also has a significant impact on economic activities [2, 3]. Billions of dollars are lost due to the treatment of injuries, loss of property, lost working hours, and high fuel consumption [4]. It is a common belief that more than 95% of accidents are attributed to human error [5]. Vehicles automation of vehicles is one of the fundamental applications of future intelligent transportation systems (ITS) to solve such problems. Vehicular ad hoc Network (VANET) has emerged mainly to improve road safety and traffic efficiency and provide passenger comfort. While the ITS focuses on providing intelligence to the roadside and vehicles systems, VANET focuses on providing communication between those systems. In VANET, vehicles are provided by sensors, actuators, computation and communication capabilities to share their observations about their state, road conditions and traffic situation in forms of periodic messages so-called Cooperative Awareness Messages (CAMs) in the European standardization framework [6]. CAMs are also called Basic Safety Messages (BSMs) in the United Sates' WAVE standards [7]. Some researchers refer to these messages as beacon messages [8], mobility data [9], or driving context information [5]. In this paper, the terms mobility information and context information are used to refer to the content of the CAMs messages. CAMs contain information about the originator, such as identification, position, time, speed and acceleration. In VANET, vehicles continuously broadcast their own context information to make neighbouring vehicles aware about the status of their surrounding vehicles to enhance drivers' perceptions about the traffic and the driving environment [10, 11]. These messages are broadcasted every 100 millisecond in a single-hop communication type with communication range up to 1 kilometre [12]. By utilizing this information vehicles perception will be extended beyond the ability of the local physical sensors such as cameras and distances sensors. Many applications and network service heavily rely on the accuracy and the reliability of such information. For example, vehicles analyse neighbouring vehicles' information and timely and autonomously detect traffic anomalies and change their behaviours accordingly, to avoid accidents and congested areas [13-15]. Routing protocol. With VANET vehicles automation can become a reality [16].

The performance of VANET's applications and services depends on the availability of accurate and recent mobility information of neighbouring vehicles [17-21]. Unfortunately, due to harsh vehicular environments, high mobility and density of vehicles, and the presence of cyber attackers, mobility information suffers from inaccuracy due to surrounding heterogeneous and dynamic noises, incompleteness due to unreliable communications, and untrustworthiness due to threats of attackers and faulty sensors [22, 23]. Besides, because VANET has a direct connection with public *safety* and *economic* activities, it may be a target of many *cyber-attackers* with high financial resources, e.g., cyberwarfare and terrorist attack. Most of the attackers in VANET exploit the mobility information to achieve their objectives due to the ease of implementation of effective attacks which have a massive impact on applications and network performance, and the difficulty of detecting such attacks. A severe adversary who has gained access to a vehicle's internal network (GPS, kinematic sensors, etc.), can manipulate the vehicles CAM messages sent to the neighbouring vehicles. For example, an attacker, which could be a malware resides in the vehicles OS, generates fake mobility patterns that simulate vehicles braking, congestions, or suddenly stopping to create false traffic anomalies. Such false mobility patterns create illusion and stimulate the applications and vehicles systems to respond to none existing event. Once the vehicles are under the attacker's control, many cyber and physical threats can be performed against the transportation systems and people lives. It is a common belief that misbehaving in terms of manipulating or misusing vehicles mobility information (the CAM messages) can disrupt the fundamental operations of any potential VANET application [24, 25].

There are many security approaches have been proposed to protect vehicles from processing false information. A detailed review of such solutions can be found in [26]. These solutions can be categorized into prevention and detection approaches. Prevention security mechanisms such as cryptographic techniques are effectively used to prevent attackers from manipulating other vehicles' information. However, cryptographic techniques are insufficient for securing VANET against manipulating own information like those who send "legitimate but false content (messages)" [27, 26]. Misbehavior Detection System (MDS) acts as a second wall of defense when the prevention mechanisms are broken to ensure the quality of the data that are shared by neighboring vehicles. Misbehavior detection in VANET has been the subject of research for many years [28, 22, 29, 26, 30-32]. Misbehavior detection solutions are categorized into two classes based on the detection target: Entity-Centric (EC) for detecting the misbehaving vehicles [28, 22, 26] and Data-Centric for detecting the false information shared among vehicles [33-35]. Data-centric misbehaviour detection have been widely suggested, due to its suitability for critical applications which instantaneous detect the attacks locally and in real time and its ability to work under privacy protected environments. In data-centric detection, vehicles evaluate the correctness of the information by investigating the consistency and the plausibility of the data that the vehicle sends [33-35]. A message is classified as false (misbehavior message) if its consistency and plausibility scores exceeded preset thresholds, and accordingly, the sender vehicle is accused as a misbehaving vehicle. Unfortunately, in VANET, the environment which faces the security solution is not stationary, and the quality of the information collected from neighboring vehicles is highly dynamic, and it is believed to have temporal and spatial correlations. However, existing misbehavior detection solutions rely on predefined and static security thresholds to assess the consistency and plausibility of the information. In doing so, high false alarms and low detection rate are expected due to the lack of consideration of the highly dynamic vehicular context. Most of the existing misbehavior detection solutions assume the availability of accurate and recent context information which is not a realistic assumption. Therefore, a constantly changing environment where mobility information has variable quality together with the absence of suitable context references lead misbehaviour detection solutions (or other applications) to be ineffective: they produce low detection rate and high rate of false alarms. Context-dependent thresholds have been reported by many researchers for future consideration as important and major elements affecting the performance of the MDS [36, 27, 37, 22]. However, at the best of our knowledge, context-aware misbehaviour detection has not been extensively studied in VANET yet.

In this paper, a context-aware misbehavior detection scheme is proposed. It aims to effectively detect the misbehavior messages through temporal and spatial analyse the uncertainties of the mobility information collected from neighboring vehicles. The proposed misbehavior detection scheme consists of four phases, acquisition, sharing, analysis, and detection. The output of each phase is used as input for the next phase. An improved Adaptive Klaman Filter algorithm for dynamic and heterogeneous noise environment is used to estimate its mobility information in the acquisition phase. Then, a driving-situational aware mobility information sharing scheme is used to effectively and efficiently share the neighboring vehicle mobility information with their uncertainties in order to provide effective quality of the shared information with efficient bandwidth utilization. In the third phase, consistency and plausibility models for both the vehicles and the context is constructed online and timely updated. The innovation error (the uncertainties) of the Kalman filter algorithm is used to build a temporal consistency model for each vehicle according to its received information using Box and Whisker Plot and used as vehicle consistency model. To reduce the false alarms, each new received message is scored according to its deviation from the representative vehicle temporal consistency model. Then, a temporal global consistency reference model is constructed using Hampel Filter utilizing the Box and Whisker Plot scores of all neighboring vehicles and used as a context reference. All models are constructed online, exploiting the temporal and spatial correlation of the innovation errors of the neighboring vehicles. Similarly, multi-faceted plausibility models are constructed using the Hampel filter for both the context and the individual vehicles utilizing the consistency reference model. Finally, the message is considered false if it deviated much from the global context reference. The results of context-aware detection model are promising to reduce the false alarms and improve the detection rate.

**Contributions.** Our contribution is a context-aware data-centric misbehaviour detection scheme (CA-DC-MDS) that utilizes the spatial and temporal correlation of the mobility information collected from neighboring vehicles to build a context reference model that is utilized to classify the Cooperative Awareness Messages (CAM) into correct or false messages. The proposed data-centric scheme aims to timely classify the messages using lightweight and unsupervised statistical techniques so it can be beneficial for real-time applications and protocols such as safety applications and routing protocols. It can be hosted in both RSU and in the vehicles OBUs for local detection to capture the attack in its initial stages. It continuously tracks the consistency and plausibility of the mobility information of the neighboring vehicles without exposing their privacy.

The rest of this paper is structured as follows. In Section 2, the related work is studied. The problem formulation and the solution concept are presented in Section 3. In Section 4, the proposed context-aware misbehavior detection model (CA-DC-MDS) is detailed. The procedure of the performance evaluation is explained in Section 5. The results and discussion are elaborated in Section 6. Finally, the conclusion and future work are given in Section 7.

## 2. Related Work

Misbehavior detection has been the subject of research for many years [28, 22, 29, 26, 30-32]. Most of the existing detection mechanisms are data-centric which aims to detect the false messages through analyzing the consistency and plausibility of the information against predefined security thresholds. In data-centric detection, vehicles evaluate the correctness of the information shared in the networks regardless of their senders [33-35]. Data-centric detection approaches are categorized into two types, as follows: event-information-based or context-information-based. Event-information-based misbehaviour is specific to covering certain types of events, such as hard breaking, stopping vehicles, congestions, and crash notification [38], which make it too application specific. Generalization of this approach has been suggested by many researchers such as in [33, 34, 39], but it is still in the level of abstraction. Context-information-based misbehaviour is the focus of this study for three main reasons: (1) by manipulating context/mobility information it is easy to perform many kinds of attacks, including detecting false event messages, (2) manipulating context information can have severe impact on people's safety, traffic efficiency and the network performance, (3) it is difficult to be detected, due to VANET's challenging VANET characteristics such as high mobility, density, and environmental noises. Misbehaving vehicles which share false context information through manipulating their own mobility information to resemble a traffic pattern event, such as hard braking, crash pattern, or congestion pattern can create illusion and cause neighbouring vehicles to accept false information [40-44]. Context-information-based misbehaviour detection has many advantages over event-information-based misbehaviour detection because it can detect the attacks in their initial stages before [27, 45, 46, 38]. Manipulating context information can trigger the applications of benign vehicles to spread false information such as Electronic Emergency Brake Light (EEBL), Slow and Suddenly Stop Vehicle message (S&SV), Post-Crash Notification (PCN), Congestion Message (CM) among many others. Such false information can have a severe impact on public safety and traffic efficiency. Detecting the validity of such information in the true time is a challenging task and still open research problem. In this study, the aim is to detect false context information before it developed to false traffic pattern and before it triggers benign vehicles to detect false traffic pattern.

Detecting false context information has been the focus of many researchers in the literature [28, 22, 29, 26, 30-32]. Two main concepts have been used for detecting false information in VANET: plausibility and consistency concepts. Plausibility-based detection uses a particular known data model of the real world for detecting implausible information [22, 30]. Messages' plausibility validation needs a set of pre-defined rules or protocol specifications which describe the physically impossible content. The plausibility model can vary from narrowly defined rules, such as violating laws of physics (e.g., violating maximum distance movement), up to rules that allow a high range of variation, such as those affected by context information accuracy. Examples of implausible content include vehicle report a speed of 700km/h, two vehicles occupying one position in the same time, a vehicle that exists in multiple positions at the same time, and transferring information beyond the communication capabilities. Such content is an obvious indication of incorrect messages. Some plausibility rules include determining the minimum and the maximum boundaries of messages' delay tolerance, velocity, positioning error, and broadcasting frequency. A major advantage of plausibility-based detection is that the message plausibility verification is simple, which makes it efficient for real-time applications. Moreover, there is no need for assumptions such as an honest majority, so that it is robust against colluding attacks.

Because position information is essential information in the context messages, Hubaux, *et al.,* [47] has proposed verifiable multilateration method which relies on Graph-Based Distance Bouncing, Time of Flight (ToF) and multilateration Technique to detect false position information. However, such a method is not practical because it needs massive roadside infrastructures. Leinmüller, *et al.,* [48] propose a model for the Maximum Communication Ranges (MCR), and the Acceptance Range Threshold (ART) as a measure of message plausibility. However, such model relies on signal strength which is unreliable in the dynamic environment due to the attenuation and the interference that could result in poor detection accuracy. Signal strength indicator (RSSI) also has been suggested as the main component to detect false information in Demirbas and Song [49] model. However, such solution uses linear attenuation model for the evaluation which assumes Gaussian and stationary positioning noises in VANET environment which does not hold.

Schmidt, *et al.,* [50] proposed a misbehaviour detection scheme which consists of several detection models to detect false information. A model for detection of Sudden Appearance Vehicles (SAV) has been proposed as an indicator of false information. The model assumes that vehicles normally appear on the edge of the communication range. It uses the signal strength to estimate the distance of the first occurring in the communication range of other vehicles that hosts the model. Another model is to observe the Minimum Moved Distance (MMD). MMD checks whether the vehicles movement distances are compatible with its speed or not. As the attacker keens to convey the false pattern, the message broadcasting rate may be put in its maximum. Hence, the scheme employs the Maximum Broadcasting Frequency (MBF) as an indication of misbehaving and possibly false information. Although such a scheme is useful for detecting false mobility messages, it lacks considering the dynamic vehicular context due to the usage of the predefined detection thresholds in the proposed detection models. It assumes the availability of accurate and reliable context messages which is not always feasible in VANET.

Bissmeyer, *et al.,* [51] devised a plausibility model for detecting false information by investigating the Overlapping Positions (OL) in the vehicles context information. A message is considered incorrect if its reported position overlapped with the location of another vehicle. This model can be useful for detecting ghost/hidden unreal vehicles. However, this assumes a static vehicular context and the harsh and unreliable communication environment have not considered in the model. Moreover, there is no way to know which of the overlapped messages is false and which one is the genuine one which may lead to increase the false alarms. Parker and Valaee [52] have suggested a detection model that ensures wither the reported position information is within road boundaries. However, the dynamicity and the harsh vehicular context have not been considered.

Stübing, *et al.,* [53] has proposed a Mobility Data Verification Scheme based on estimation theory. Kalman filter algorithm is used to track the consistency of the neighboring vehicle's movement. Kalman Filter checks whether the new arrival CAMs messages are consistent with their preceding. A predefined threshold has been used to decide whether the new arrivmobility information is consistent or not. Although such an approach can be useful, the use of a static threshold in VANET is the main limitation. It was assumed that the positioning noises are stationary and Gaussian which is an unrealistic assumption for VANET [54]. Moreover, the scheme assumes the ideal communication and no loss of the context information. Such a hypothesis cannot be satisfied due to the high mobility, density and the requirement of broadcasting at a high rate. In line with the use of state theory-based solution, Bissmeyer, *et al.,* [45] proposed a Mobility Data Verification Scheme based on a particle filter. Unlike Kalman filter, particle filter could be an effective solution for nonlinear problems, however, the large number of particles used in high vehicles density and mobility are not feasible for real-time applications. Bissmeyer, *et al.,* [22] has proposed a multifaceted Message Rating Scheme to detect false context information. The scheme employs the Kalman-filter based scheme proposed by Stübing, *et al.,* [53] to rate message temporal consistency. Then, it rates the message plausibility using the overlapping model that was proposed by Bissmeyer, *et al.,* [51] and the ranging based models that were proposed by Schmidt, *et al.,* [50]. The proposed scheme lacks considering the vehicular context characteristics such as high dynamic context namely the unreliable communication channels and the heterogeneous noise environment.

Root-Cause-Tree [55] apply several machine learning algorithms to classify the messages into two classes: true or false. It is being reported that the Random Forest algorithm has achieved the best accuracy among the tested algorithms. However, the datasets that have been used for training and testing generated under controlled conditions and a specific scenario. The dynamic vehicular context and harsh environment have not considered thus such finding cannot be generalized. Zaidi, *et al.,* [38] has proposed a traffic model for detecting false traffic information. The proposed statistical model relies on macroscopic traffic variables such as traffic flow and traffic density to detect abnormal traffic patterns. Unlike the previous models [50, 51, 53, 55, 22] that are relying on the microscopic traffic variables such as vehicles movement traces. This model can be effective to detect abnormal traffic behavior for non-real time applications like congestion avoidance and traffic control. The use of macroscopic traffic variables makes the model rely on coarse-granularity context information, i.e. the attack cannot be detected in its initial stages. It is only detects the attacks when it developed to false traffic patterns such as congestions, hard braking, and etcetera.

In summary, most of the existing context-based misbehavior detection solutions are designed to detect false context information under controlled conditions and stationary environment [56, 57, 46, 58]. However, such solutions are unaware of the highly dynamic vehicular context and harsh environment. The use of predefined security thresholds is the major drawbacks in the high dynamic and harsh environment. The lack of consideration the heterogeneous noises the influence the context-information during to the movement of vehicles in harsh environments and the lack of consideration of the unreliable communication in VANET due to the high density and mobility, such solutions tend to produce high false alarms and low detection accuracy [59, 16, 26].

## 3. Problem Formulation and Solution Concept

The use of static security thresholds is the main limitation of the existing solutions. Attackers can easily inject false context information under these thresholds, so that it is difficult to be detected [16]. Plausibility and consistency thresholds should be able to cope with the context change to increase the detection accuracy and decrease false alarms [59, 60]. Fig. 1 shows four context situations in which data-centric misbehaviour detection is ineffective. The uncertainties of the context information collected from neighboring vehicles are dynamic. The dotted elliptical shapes represent the perimeters of the normal situation while the solid line circles represent the static security threshold (the predefined threshold as used by the existing MDS solutions). It can be noticed that the predefined context is ineffective in terms of the detection rate (see Figure 1 (A) and (B)) and the false positive (see Figure 1 (C) and (D)).

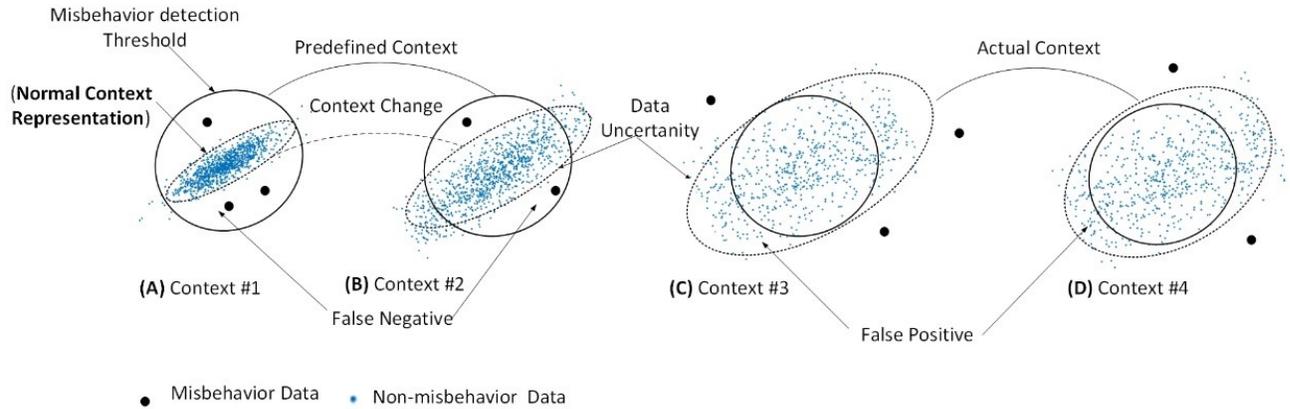

**Fig** Error! No text of specified style in document.**:** Dynamic Context with Fixed Security Thresholds

In this study, a context-aware data-centric misbehaviour detection schemes (CA-DC-MDS) is proposed as an improvement to the context-unaware schemes that was proposed in [50, 51, 53, 55, 22]. Context-dependent thresholds have been reported by many researchers for future consideration as important and major elements affecting the performance of the MDS [36, 27, 37, 22]. For example, Zhang, (2012) suggested that evaluating the information received from other vehicles should be made in a particular context. However, to the best of our knowledge, no context-aware misbehaviour detection system is available. Similarly, the dynamic uncertainty of the information has also been reported in the literature as challenging, but, to the best of our knowledge, there are no studies have addressed this issue [46, 22]. The temporal and spatial correlation of the context information is utilized to construct the consistency reference model namely the context reference model in term of the message consistency. The proposed scheme uses the consistency and plausibility models that are built online and continuously updated to classify the messages. While previous consistency models rely on the self-consistency, the proposed model extends this concept to include the spatial consistency i.e. the consistency with the message of the neighboring vehicles. As neighboring vehicles are exposed to similar traffic and environmental conditions, such procedure is essential to reduce false alarms. In the same way, the plausibility reference models are constructed online and timely updated utilizing the consistency reference model.

## 4. The Proposed Context-Aware Misbehavior Detection Scheme

The goal of the proposed CA-DC-MDS scheme is to detect false CAMs messages effectively. The design of the CA-DC-MDS scheme consists of four phases: context acquisition, context sharing, context assessment, and misbehavior detection phase (see Fig. 2). Hence context-information is acquired in a highly dynamic and harsh environment and it shared in an unreliable communication channel, the uncertainty of the context information is high. Accordingly, the uncertainty of the information should be taken into consideration as an important context feature that influences the performance of the decision making in VANTE. Thus, in the first phase, an acquisition algorithm which was developed by the authors of this paper is used to acquire the context information. It can accurately estimate both the accuracy and the uncertainty of the vehicles mobility information in real-time. Because each vehicle should share the acquired information of their own with the neighboring vehicles, in the second phase, a broadcasting rate that can preserve the accuracy while maintaining low broadcasting rate should be used in VANET communication environment. Thus, recent context information of all neighboring vehicles is available to each vehicle. In the third phase, each vehicle extracts the consistency and plausibility features from the context information and construct the context representation reference model in terms of consistency and plausibility of the information. In the fourth phase, the newly received messages are assessed using the assessment models and compared to the recent context reference. If the score of the newly received CAM message deviates much from the context reference, then the message is considered misbehavior (or false).

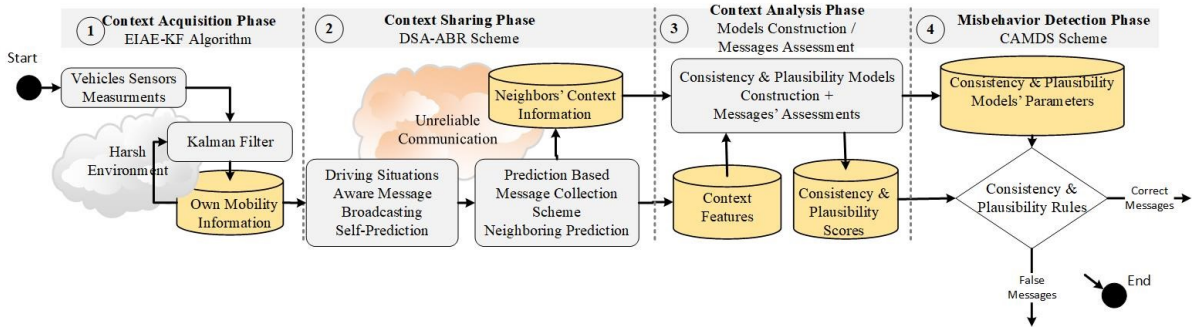

**Fig. 2:** 4-Phases Context-Aware MDS

A. Context Acquisition Phase

Because context information is measured under harsh and dynamic noise environment, the uncertainty of the information is high. Positioning information, which is the most important element in the mobility message, is the most element affected by VANET harsh environment. Position sensors are susceptible to heterogeneous and dynamic noises which render the uncertainty of the position information is high. Despite there are many algorithms have been proposed to improve the positioning accuracy in VANET environment, the heterogeneity of the noises and the high dynamics of VANET make the accuracy of the poisoning algorithms is dynamic and unreliable. Therefore, the positioning algorithm that is robust to the heterogeneous environment, able to cope with the highly dynamic nature of VANET, and can estimate the uncertainty of the acquired context information should be used. Among these algorithms, the positioning algorithm that has been proposed by the author of this paper is selected namely the Enhanced Innovation Adaptive Estimation based Kalman Filter (EIAE-KF) algorithm [54]. The EIAE-KF algorithm uses the innovation errors of the Kalman Filter to estimate the potential noises that impact the positioning sensors and accordingly update the uncertainty parameters of the Kalman filter algorithm in real-time. In doing so, the Kalman filter becomes more accurate, and the uncertainty of the information can be accurately estimated.

B. Context Sharing Phase

In VANET, vehicles periodically broadcast their mobility information at a high rate and single hop communication mode. Each vehicle collects the context information from the neighbouring vehicles to be aware of the mobility status of the neighboring vehicles. However, the high vehicles' density and mobility make the communication channel congests and thus the loss of the information is a common problem in VANET. In such a situation, the context information becomes incomplete and unreliable. Unfortunately, many security solutions have overlooked this issue and assume an ideal communication channel which is not a realistic assumption. Thus, selection broadcasting scheme is important for the design of both applications and security services. The most viable one that can reduce the broadcasting rate while preserving the quality of the context information. In this study, a driving situation aware of broadcasting rate scheme DSA-ABR is used. DSA-ABR can reduce the broadcasting rate while accurately predicting the lost or omitted mobility messages. In DSA-ABR, each vehicle individually decides to either broadcast or omit the CAM messages according to the movement patterns. If vehicles movement is stable, a lower broadcasting rate is used. In contrast, if the mobility pattern changed vehicles increase the broadcasting rate. Because not all vehicles run into a critical driving situation at the same time, the computation of the communication channel will be low. DSA-ABR scheme consists of two algorithms: self-predictor and neighboring predictor algorithm. The self-prediction algorithm is used by sender vehicles to ensure wither the own context information can be predicted by neighboring vehicles or not given last broadcasted message or not. If the prediction error exceeded a specified threshold, the message is broadcasted; otherwise, the message is omitted. In such a scheme a message is broadcasted either if a significant change happened from its preceding, i.e., if the individual driving situation has been changed or a time threshold is passed. The uncertainty of the information also shared between the vehicles to improve the prediction quality of the neighbouring data. Hence, the broadcasted message format includes the position, speed, acceleration, and uncertainties as follows $m(ID, Position, Speed, Acceleration, Unceranity)$. In the neighbor-predictor algorithm, each vehicle tries to predict fine-grained information (each 100ms) utilizing the last received information from the sender. The neighboring predictor algorithm works as a multi-target tracking algorithm. In doing so the neighbouring vehicles will be continuously aware of the mobility of the neighbouring vehicles and also sure about the quality of the context information.

C. Context Analysis Phase

This phase aims to generate a context reference model in terms of the consistency and plausibility of the context information collected from the previous phase.

a. *Context Information Consistency Assessment Model*

This section aims at assessing the consistency of the context information messages as well as constructing a context reference in terms of the consistency of the information. This aim is achieved by two steps as follows.

1. Temporal Consistency Assessment Model

The first stage includes deriving the representative consistency features from context information that was collected through the broadcasting scheme that implements Kalman Filter to track the messages of all neighbouring vehicles. The innovation sequences describe the discrepancy between the neighbour prediction algorithm and the reported context information (CAMs) from each neighbours.

$$z_{(x,y)k} = y^{NP}_{(x,y)k} - \check{y}^{NP}_{(x,y)k|k-1} \quad (1)$$

where $z_{(x,y)k}$ is the innovation vector contains the latitude $x$ and longitude $y$ of the neighboring vehicles at time epoch $k$. $y^{NP}_{(x,y)k}$ is a vector that contains the last received latitude $x$ and longitude $y$ values of the neighbouring vehicles. $\check{y}^{NP}_{(x,y)k|k-1}$ is a vector that contains the predicted latitude $x$ and longitude $y$ values of the neighbouring vehicles by Kalman Filter prediction model. $z_{(x,y)k}$ is used as input for the temporal vehicle consistency model. The second stage is the temporal analysis of context information. The Box-and-Whisker plot algorithm is used to analysis of the temporal deviation of innovation error $z^i_{(x,y)k}$ values for a neighbouring vehicle $i$ from its preceding values (innovation sequence $z^i_{(x,y)k:k-w}$). Box-plot is a non-parametric statistical tool that can summarize the time series variable without the need to know its underline distribution. Each received context information message is assessed according to its temporal summary generated by Box-Plot. The temporal consistency summary consists of four main elements which are the temporal median $\mu_k$, entire quartile range $IQR$, the upper adjacent value (Upper Limit $UL_k$), and the lower adjacent value (Lower Limit $LL_k$) of the innovation sequence of each neighbouring vehicles. Thus, the temporal consistency summary can be represented by quadruple elements denoted by $TS_{k(i)}(\mu_k, IQR, LL_k\ UL_k)$. The temporal consistency $TS_{k(i)}$ for a vehicle $i$ at every time epoch can be written as follows.

$$TS_{k(i)} = \begin{cases} \mu_k = (Q_1 + Q_3)/2 \\ IQR = (Q_3 - Q_1) \\ UL_k = Q_3 + 1.5\ IQR \\ LL_k = Q_3 - 1.5\ IQR \end{cases} \quad (2)$$

where $Q_1$, and $Q_3$ are the first and third quartile of the innovation sequence, respectively. Accordingly, the temporal consistency score as follows.

$$TCS_{k(i)} = max(UL_{k(i)}, \check{z}_{(x,y)k(i)}) \quad (3)$$

where $TS_{k(i)}$ is the temporal score a message received from vehicle $i$ at any time epoch $k$ is set to the upper adjacent value $UL_{k(i)}$ if $z_{(x,y)k}$ is located under the upper adjacent value $UL_{k(i)}$ otherwise $TS_{k(i)}$ is assigned to $\check{z}_{(x,y)k(i)}$. Because the temporal consistency score is context dependant, relying on the temporal deviation for misbehaviour detection will result in a considerable amount of false alarms. To tackle such a problem, it is necessary to differentiate whether the temporal deviation of temporal scores is due to misbehaviour or otherwise. To this end, a spatial filter is used to detect the outlier where temporal scores of misbehaving vehicles are located far from the major temporal scores of the messages received from neighbouring vehicles.

2. Spatial Consistency Assessment Model

The temporal scores $TCS_k$ of all neighbouring vehicles are used to generate a spatial consistency model that is used as consistency reference for spatial assessment. An essential assumption is made that the existence of an honest majority in the neighbouring vehicles. This assumption is reasonable in VANET and has been widely used in the literature [33, 61]. Hampel filter is used to generate the spatial consistency reference model ($SCRM$) to assess the spatial deviation of $TCS_{k(i)}$ of a neighbouring vehicle from its neighbouring vehicles' $TCS_{k(j)}$. Hampel filter is a non-parametric statistical filter which is used to estimate the mean and standard deviation as a robust alternative to outlier sensitive z-score ($z = \frac{x_t - \mu}{\sigma}$). It replaces the arithmetic mean $\mu$, and standard deviation $\sigma$ by the median and median absolute deviation ($MAD$) respectively. The arithmetic mean and standard deviation deviate significantly if the data contain outliers, thus they are ineffective in the presence of misbehaviour data. The algorithm in Fig. 3 illustrates the procedures to compute vehicles' spatial consistency scores $SS_{k(N,W)}$ using Hampel filter.

---

**Algorithm 1: Hampel-Filter Based Spatial Consistency Assessment Model**

**Input:** $TCS_{k(n \times w)}$    // the temporal consistency score matrix for last received messages $n$ at the time epoch $k$ with time window length $w$

**Output:** $CICAM_{Model}$ //The spatial consistency reference model ($CICAM$)
  $SCS_k$      //The spatial consistency score vector for all received messages in the time epoch $k$

1:  Set $CICAM_{Model} = \begin{cases} \emptyset_k = \mu_k = median(TS_{k(n \times w)}) \\ \delta_k = \sigma_k = 1.4826 \times median\{|TS_{k(n \times w)} - \emptyset_k|\} \\ HUB_k = \emptyset_k + \beta \times \delta_k \\ HLB_k = \emptyset_k - \beta \times \delta_k \end{cases}$

2:  Calculate $SS_{k(i)} = \left|\frac{TS_{k(i)} - \emptyset_k}{\delta_k}\right|$

3:  Update $SS_k$ Vector
4:  Return $CICAM_{Model}, SS_k$

---

**Fig. 3:** Hampel filter-based spatial Consistency Assessment Model

| Symbol | Description |
|---|---|
| $\emptyset_k$ | The median at time epoch $k$ |
| $\delta_k$ | The Hampel filter based median absolute deviation (MAD) |
| $HUB_k$ | The Hampel upper boundary of the CICAM model |
| $HLB_k$ | The Hampel lower boundary of the CICAM model |
| $\beta$ | A threshold that represents the accepted deviation in the context information. |
| $SS_{k(i)}$ | The spatial score of vehicle $i$ at time epoch $k$. |

As shown in Fig. 3, the outputs of this stage are the $CICAM_{Model}$ which contains the spatial consistency reference of vehicular context and $SS_k$ vector which contains the spatial score of each neighbouring vehicles. $CICAM_{Model}$ is used as a reference model to represent the consistency of the context information in a particular context. Thus, it is used for both assessing the spatial consistency of context information messages and detecting the misbehaviour information in the following sections.

b. *Context Information Plausibility Assessment Models*

There are two context dependent models that have been used in the literature which need to be modified: the overlap-based plausibility assessment, and the communication range-based model.

1. Overlap-based Plausibility Assessment Model (OLPAM)

   OLPAM aims to assess the plausibility of the information in terms of overlapped occupation area of neighbouring vehicles. In normal status, two vehicles cannot occupy the same area at the same time. The accuracy of this model depends on the quality of the context information. As the accuracy of the context information is context dependent, $OLPAM_{model}$ should shrink the occupation area of a vehicle according to the uncertainty of context information. Therefore, the maximum boundary of consistency of the context information ($HUB_k$) of $CICAM_{model}$ is used in this model. In doing so, the model accurately distinguishes between false information and inaccurate information due to the context. To construct the $OLPAM_{model}$ four input are used as follows: the position information in the context information, vehicles dimensions (length, and width), and $CICAM_{model}$. The output of this module is $OLPAM_{model}$ and the overlapping plausibility score $OLPS$. Fig 4 (a), (b), and (c) show the concept of the proposed overlap-based plausibility assessment model.

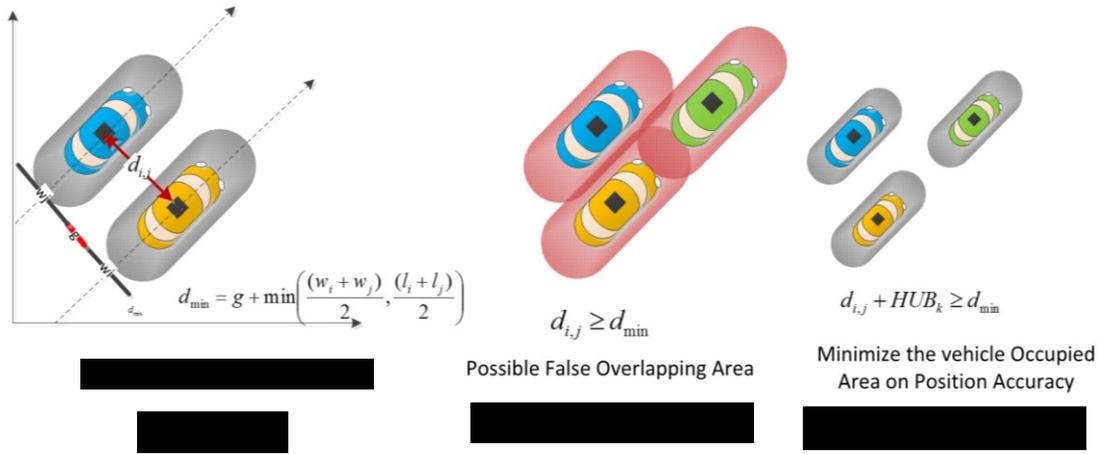

**Fig. 4:** Context-Aware Overlap Detection Concept

Fig. 4 (a) shows the concept of overlapping detection in which the distance $d_{(i,j)}$ between the positions of two vehicles $i$ and $j$ should by greater than a threshold $d_{min}$. This threshold ($d_{min}$) is the minimum acceptable distance between two vehicles. The overlapping plausibility assessment model can be explained as follows. Firstly, each vehicle is represented by rectangular while the certainty of the position information is represented by an oval shape. The centre of the rectangle can be at any position on the uncertainty oval circumference. Thus, vehicles rectangle shape is enlarged as shown in Fig. 4 (a). If the uncertainty of vehicles position is high, then the overlapping is highly possible as shown in Fig. 4 (b). This results in high false alarms in many situations. Existing plausibility models reduce false alarms by enlarging the uncertainty threshold. Such remedies render the model ineffective because it decreases the model sensitivity. In this study, to reduce false alarms and keep the detection rate at a high level, a dynamic uncertainty threshold is used. During high uncertainty, the rectangle that represents the vehicle is shrunken (See Fig. 4 (c)). The rectangle shape is shrunken or enlarged according to the uncertainty of the context information. Thus, a message is consider overlapped if the following equation is satisfied.

$$OL_{i,j} = \begin{cases} 1 & |d_{(i,j)} + HUB_k| < d_{min(i,j)} \text{ //Overlap} \\ 0 & |d_{(i,j)} + HUB_k| \geq d_{min(i,j)} \text{ //noneOverlap} \end{cases} \quad (4)$$

where $OL_{i,j}$ is the overlapping status between vehicle $i$ and $j$. Thus, the overlapping score $OLPS_{i(k)}$ of message received from vehicle $i$ at time epoch $k$ is the number of time it overlaps with any message received from neighbouring vehicles at the same (see Eq. 5).

$$OLPS_{i(k)} = \sum OLP_{i \text{ with any } (k)} \quad (5)$$

1. **Communication-Range-based Plausibility Assessment Model (CRPAM)**

Fig. 5 shows example of communication range based plausibility detection concept. Normal vehicles are firstly appearing on the edge of the communication range while misbehaving vehicles called Sudden Appearance Violation (SAV) attackers may appear directly in close to victims. In addition, normal vehicles send within communication range while the misbehaving vehicles called maximum communication range violation (MCRV) attackers may manipulate transmission power and appear outside the default communication range.

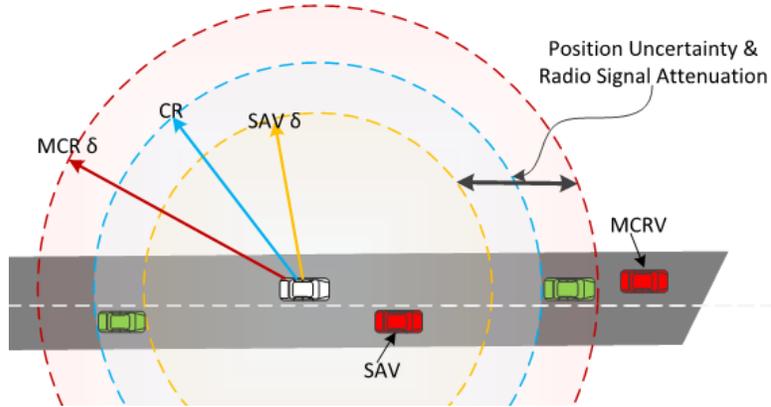

**Fig. 5:** Communication range bounds

A message is considered false if it violated either the minimum appearance threshold or maximum appearance threshold. Existing plausibility models use a pre-set thresholds. However, these thresholds are context dependent both the signal strength and the position information is context dependent. Thus, relying on a pre-set and static thresholds increases the false alarms rate and/or decreases the detection rate. In this study, each vehicle constructs a model from the distances of occurring of the neighboring vehicles. To reduce the false positive rate, the occurrence distances are modified according to the uncertainty of the information using the maximum boundary of consistency of the context information ($HUB_k$) of $CICAM_{model}$. Hampel filter upper bound is used to represent the maximum communication range while Hampel filter lower bound is used to represent the threshold for the accepted sudden appearance vehicle (SAV).

$$SAV_j = \begin{cases} 1 & |d_{(i,j)} - HUB_k| < SAVT \\ 0 & \text{otherwise} \end{cases}$$

$$MCRV_j = \begin{cases} 1 & |d_{(i,j)} + HUB_k| < MCRT \\ 0 & \text{otherwise} \end{cases}$$

D. The Misbehavior Detection Phase

In the previous three sections, three modules were developed namely $CICAM$, $OLPAM$, and $CRPAM$. Each module is responsible to construct a spatial reference model online and continuously updated. Each module also assesses the consistency and plausibility of the newly arrived context information messages.

Misbehaviour decision fusion module aims to classify the arrived context information messages (CAM) into two classes: false and true messages. Four context-adaptive rules are proposed, namely consistency-based rule, overlapping-based rule, maximum communication range-based rule, and sudden appearance-based rule which can be described as follows.

1. **Rule 1:** consistency rule: a message is misbehaviour if the consistency score of the message exceeds the Hampel filter upper bound HUB of the CICAM model ($HLB_k \leq CS_{k(i)} \leq HUB_k$).
2. **Rule 2:** overlap rule: a message is misbehaviour if the overlap score of the message ($OLPS_{k(i)}$) is located between overlap lower boundary $OLB_k$ and overlapping upper boundary $OUB_k$ of the $OLPAM_{model}$ ($OLB_k \leq OLPS_{k(i)} \leq OUB_k$).
3. **Rule 3:** maximum communication range rule: a message is misbehaviour if the distance between current vehicle and the sender namely $FOPS$ score is exceed the higher communication range boundary $MCRT_k$ of the $CRPAM_{model}$ ($CR_{k(i)} > MCRT_k$).
4. **Rule 4:** sudden appearance rule: this rule is applied when a message received from newly vehicles which is entered the communication area of subject vehicle. A newly arrived message is misbehaviour if the distance between subject vehicle and the position claimed in the context information message is lower than lower appearance range boundary $SAVT_k$ of the $CRPAM_{model}$ ($CRFO > SAVT_k$).

These rules are checked for each arrived message. If the rule matched the message is pass to the second rule. Fig. 6 illustrates the flowchart of the proposed context-aware data-centric misbehaviour detection scheme (CAMDS). If one of the previous consistency or plausibility models gives positive result, then the corresponding context information is discarded.

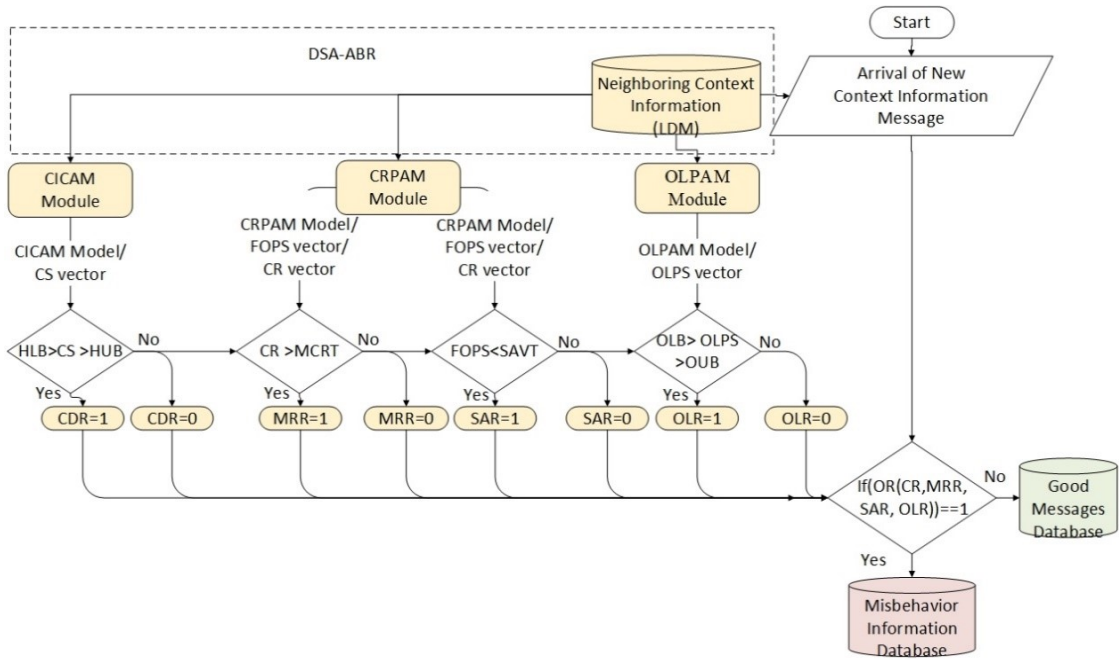

**Fig. 6:** The proposed CAMDS scheme

## 5. Performance Evaluation

A. Experimental Setup

To evaluate the proposed CAMDS, extensive simulation have been used. This includes the traffic datasets used and the procedures used for data preprocessing, selecting data samples, environmental noise injection, communication simulation and misbehavior simulation which are common procedures for evaluating misbehavior detection solutions as performed in [46, 22]. To perform the simulation, Matlab, which is a well-known simulation environment in the automotive domain, has been used for simulating the environmental noises, communication channel, and misbehaving vehicles. Matlab tools have been used by many researchers to evaluate the misbehavior detection such as in [46, 22, 62-64].

*1.1. DataSets Source and Preprocessing*

A real-world traffic dataset which contains vehicles trajectories recorded each 100ms was used namely the Next Generation Simulation (NGSIM) Dataset. NGSIM was generated by the United States Department of Transportation (US DOT) Federal Highway Administration (FHWA) [65]. It represents the ground truth information of neighbouring vehicles' trajectories [65, 66]. Similar to the approach taken by [46], instead of using the common traffic scenarios such as city and highway scenarios which are limited to low dynamic in their nature (scenarios without highly dynamic changes in vehicles driving behaviour) such as urban or rural scenarios, this study used high dynamic scenarios from NGSIM dataset. . The datasets contain many traffic scenarios in which the density and the speed of the vehicles vary. This makes the datasets be able to represent the context in terms of dynamic traffic status.

To ensure that the evaluation has considered all types of driver behaviours, the dataset is categorized into four different clusters, based on driver behaviour. For each vehicle, three features are selected to represent driver behaviour, namely: time headway, space headway and lane changing ratio. The selected features are aggregated by finding their means and variance and using these as input for the K-Means clustering algorithm [67]. These clusters describe four types of driving regime, specifically: free flowing, random flowing, car flowing, and lane changing behaviour. The purpose of this categorization is to ensure that the vehicle behaviours have no influence on the performance of the proposed scheme.

*1.2. Simulation of Environmental Noises*

Various types of environmental noise have been injected to the vehicle trajectories (NGSIM data) to represent VANET harsh environment. A combination of stationary white noises with zero mean, non-stationary white noise with time-varying variance , and correlated noises have been reported by many researchers in VANET context acquisition and used to simulate the dynamic and heterogeneous noise environmental noise in VANET [54]. Noise injection is common procedure to simulate measurement noises in VANET environment [52, 68, 20]. The noise types, noise scenarios, and their simulation have been reported in our previous publication [54]. The acquisition algorithm presented in [54] has been used to acquire own context information in each vehicle.

*1.3. Simulation of Message Losses*

Nine communication scenarios were simulated. In each simulation, the message arrival rate is modelled as Poisson distribution with 9 different message arrival property range from (1 – to 0.03 within each 100ms from each neighbouring vehicle). These nine scenarios represents different communication status in VANET. Each vehicle uses the broadcasting scheme presented in [69] to efficiently share own information and effectively collect the context information of the neighboring vehicles. Table 1 shows the dataset samples that were used to evaluate and validate the proposed CAMDS scheme. The dataset samples were extracted from 15 vehicles which were selected randomly from different driving regime. In addition to one dataset which was extracted from one simulated RSU. Each dataset has been replayed under 9 communication scenario. Thus, a total of 135 experiments have been performed.

**Table 1:** NGSIM Dataset selected samples

| Dataset | Vehicle Id | Vehicle Regime | Average Speed (m/s) | Duration (s) | Dataset Size | Total Neighbours |
|---|---|---|---|---|---|---|
| DS1 | 13 | Free-Flow | 16.8 | 94.8 | 113,258 | 177 |
| DS2 | 252 | Free-Flow | 23.3 | 55.8 | 155,908 | 255 |
| DS3 | 455 | Free-Flow | 21.6 | 60.3 | 145,904 | 260 |
| DS4 | 2280 | Free-Flow | 24.0 | 54.1 | 197,511 | 270 |
| DS5 | 5 | Lane-Change | 22.4 | 70.2 | 80,568 | 119 |

| Dataset | Vehicle Id | Vehicle Regime | Average Speed (m/s) | Duration (s) | Dataset Size | Total Neighbours |
|---|---|---|---|---|---|---|
| DS6 | 1133 | Lane-Change | 31.7 | 39.3 | 107,565 | 214 |
| DS7 | 1687 | Lane-Change | 22.4 | 76.5 | 110,051 | 314 |
| DS8 | 1 | Lane-Change | 18.0 | 88.4 | 88,971 | 134 |
| DS9 | 268 | Flowing-Mode | 26.2 | 58.5 | 156,941 | 255 |
| DS10 | 1066 | Flowing-Mode | 33.0 | 47.5 | 111,305 | 225 |
| DS11 | 1964 | Flowing-Mode | 21.5 | 72.9 | 223,211 | 317 |
| DS12 | 7 | Flowing-Mode | 22.5 | 71.1 | 85,244 | 127 |
| DS13 | 1593 | Flowing-Mode | 21.1 | 74.2 | 186,260 | 294 |
| DS14 | 2885 | Random-Flow | 16.5 | 94.5 | 150,127 | 200 |
| DS15 | 1899 | Random-Flow | 19.9 | 78.8 | 231,867 | 331 |
| DS16 | RSU | Mixed | 28.0 | 57.3 | 479,823 | 284 |

*1.4. Simulation of false Information Attack*

Due the absence of ground truth labelled dataset for evaluating misbehaviour detection systems in vehicular network, simulating the misbehaviour actions is common evaluation procedure. Two types of misbehaviours are simulated: faulty vehicles and attackers. Three types of faults were injected in the datasets spikes, noises and constant type attacks. The first two types of faults were detected and corrected before misbehaviour detection is started through the EIAE-KF algorithm as they were considered as inconsistent context information during acquisition phase. Meanwhile, the constant faults are detected by the proposed CA-DC-MDS. The second type of misbehaviour is attacker actions which can be grouped into two types: basic attack and sophisticated attack. Basic attack such as positioning noises, position jumping, message suppression attack, cheating with context information attack, sudden position jumping, and random jumping based on works introduced in [70, 71]. The sophisticated attack is introduced by the authors of this paper. The attackers of this type aware about the context, thus they perform incremental jumps to perform attack such as illusion attack that was described in [40].

## 6. Results and Discussion

This section evaluate the performance of the CA-DC-MDS scheme in terms of detection accuracy, false positive rate (FPR), detection rate (DR), and F-measure. Because the attacker data can be very few comparing with normal data then F-Measure is consider as suitable evaluation matric because it does not take the true negative into account [35]. The proposed CA-DC-MDM is compared with the scheme proposed by Stübing, *et al.,* [53]. This scheme consists of many plausibility and consistency models which was developed and improved by Firl, *et al.,* [46], Jaeger, *et al.,* [37], Leinmüller, *et al.,* [48], Schmidt, *et al.,* [50]. It is common to evaluate the data-centric misbehaviour detection using the normal dataset (before simulating the attacker actions) in terms of false positive rate (FBR) in order to evaluate its effectiveness under different communication scenarios [53, 46, 22]. Fig. 7 illustrates a comparison between CA-DC-MDS and the baseline MDS in terms of FPR using the normal dataset

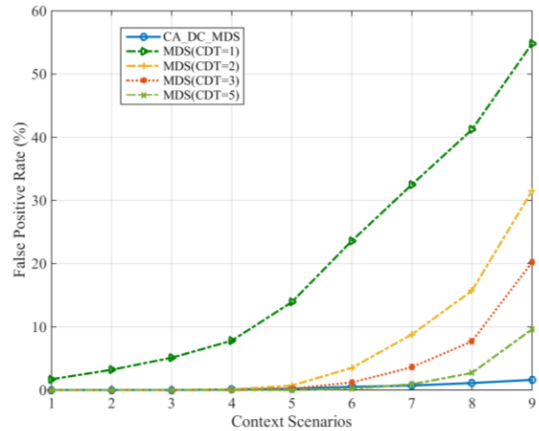

**Fig. 7:** Dataset Collection from V2X Architecture

In Fig. 7, X-axis represents the nine studied scenarios while Y-axis represents the average FPR for the 15 dataset which were extracted from different vehicles. The result shows that the performance of the baseline scheme depends on the scenario and the selected threshold. The false positive is increased dramatically until it reaches 54.8%. The increase depends on the detection threshold, the smaller of the consistency detection threshold leads to more false positive. Meanwhile, the proposed CA-DC-MDS has lower FPR which indicates the robustness of the CA-DC-MDS under highly dynamic scenarios.

Fig. 8 (a), (b), (c), and (d) depict performance comparison between CA-DC-MDS and the baseline MDS in terms of the accuracy, FPR, DR, and F-Measure, respectively using datasets containing misbehaviour information. Fig. 8 (a) shows the comparison accuracy of the proposed CA-DC-MDS to the baseline MDS. As can be seen in the Figures 8 (a), the accuracy of the proposed CA-DC-MDS is high and consistent in all scenarios including when the communication is highly unreliable. In contrast, the accuracy of the baseline scheme MDS is rapidly decrease as message loss increases. In addition, when the consistency detection threshold (CDT) is smaller, the detection accuracy becomes worst. Fig. 8 (b), and (c) compare false positive rate (FPR) and the detection rate (DR) respectively. In terms of FPR, the CA-DC-MDS achieved low false alarms with slight increase when messages loss ratio increase. Meanwhile, the FPR of the baseline MDS is rapidly increased as consistency threshold is decreases. In terms of detection rate, CA-DC-MDS has high DR which always higher than 95.8% in all studied scenarios, while the detection rate (DR) of the baseline MDS is rapidly decreased as consistency threshold is increased. This implies that the baseline MDS trade-off between the detection rate and the false positive rate which is the main drawback of the baseline model which the proposed CA-DC-MDS scheme has solved.

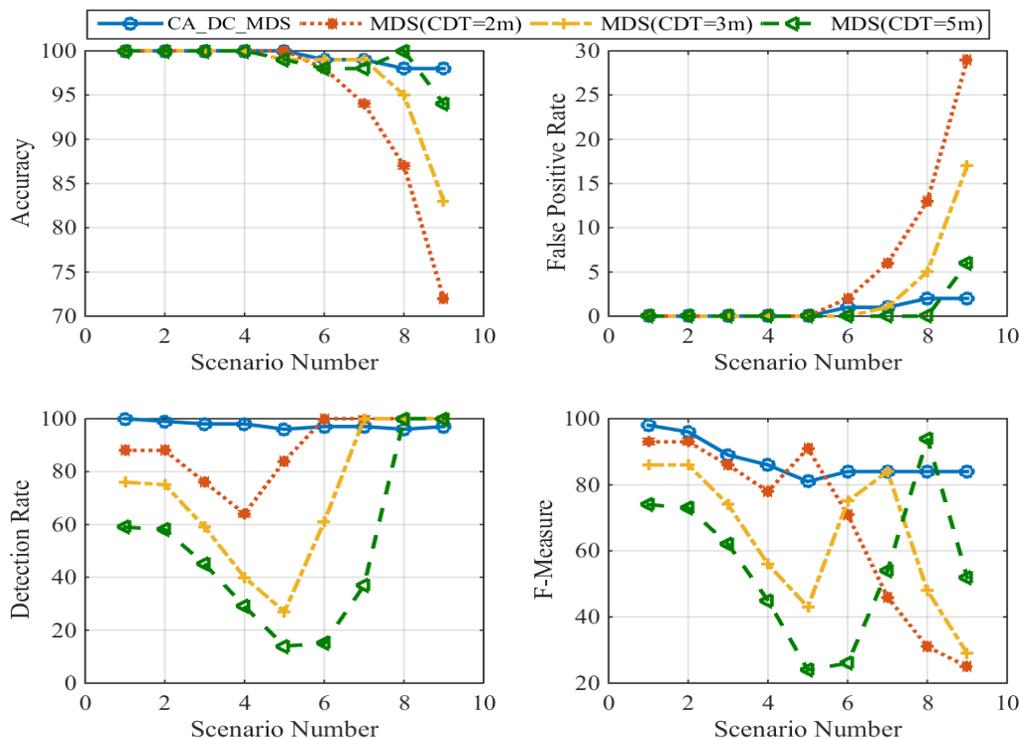

**Fig. 8:** Comparison between CA-DC-MDS and Baseline MDS

Fig. 8 (d) illustrates the overall performance of the proposed detection model in terms of balanced F-measure. CA-DC-MDS has high F-measure value with slightly dropping when communication loss increases. However, the F-measure become more consistent higher than 80% when the communication loss become worst. The reason for decreasing the F-measure is referred to the reduction of the detection rate due to increase the uncertainty of the information which makes the distinction between the normal and false information is difficult. The reason for stability higher than 80% is due to the loss of information of both normal and false which case increase of the prediction error in the prediction algorithm during the context data collection phase. When the prediction error increases the uncertainty of the information also increases, CA-DS-MDS increases the value of the threshold. Thus, more stability in the performance is achieved. In contrast, the fluctuant behaviour of the baseline MDS scheme is due to fixed threshold which is not consistent with the context. The F-measure become higher when the consistency between the static threshold and the context high, while it is drops to low performance when the quality of the context information become low. This implies the effectiveness of the proposed scheme comparing by the existing MDS.

Since the assumption is honest majority honest, the boundaries of the consistency and plausibility model could be influenced by the amount of misbehaving data. Therefore, an additional experiments were conducted in order to study the impacts of increasing misbehaving vehicles on the scheme performance. Four percentages of misbehaving vehicles were experimented as follows: 10%, 20%, 30% and 40%. Fig. 9 illustrates the results of the experiments. As shown in Figure 6.16 (a), the misbehaving ratio has a minor effect on the accuracy of the CA-DC-MDS when the communication loss is greater than 90% while the accuracy is degraded under 90% message loss. It is clear that as the misbehaving vehicles increase the accuracy is decreased. The FPR is slightly increased approaching 1% when the message loss less than 90% while it is increasing until 0% under 90% message loss scenarios. In terms of DR, CA-DC-MDS shows that the detection rate is slightly decreased when the misbehaving vehicles ratio increased. This implies that the amount of misbehaving slightly influenced the detection thresholds.

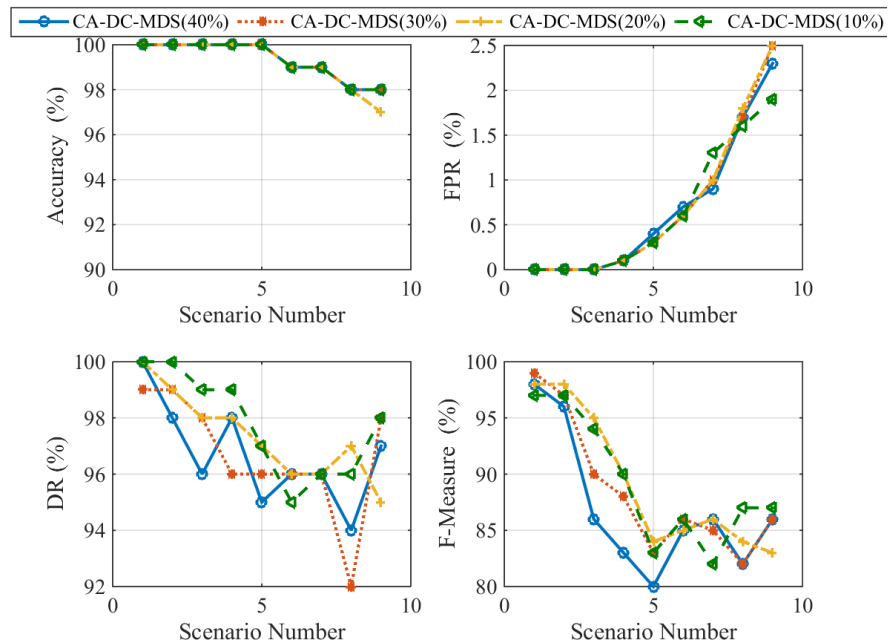

**Fig. 8:** Influence of number of Misbehaving Vehicles on the CA-DC-MDS Effectiveness

It can be noticed that the increasing number of misbehaving vehicles has to some extent has low effect on the performance of the proposed CA-DC-MDS. This positive property of the proposed scheme made the increasing of the misbehaving vehicles percentage lead to slightly increase the value of the boundaries of the spatial consistency. This is because the use of outlier robust Hampel filter which employs the median instead of the arithmetic mean to centre the data. In addition, vehicles are aware of the inconsistencies of the information so that the misbehaving vehicles will not achieve the attack objectives as the spatial consistency is increased. Therefore, there is no benefit will be achieved even if attackers colluded to perform an attack. Due to the randomness of both misbehaving vehicles behaviour and driving regime of the misbehaving vehicles, the context change is not necessary linear this interpret the fluctuant behaviour in the Fig. 8 spatially in the detection rate. Therefore, it can be concluded that the proposed model stay effective even if under high percentage of misbehaving vehicles reach to 40%. In conclusion, CA-DC-MDS can effectively perform in presence high ratio of misbehaving vehicles up to 40%.

## 7. Conclusion

Context-based data-centric misbehaviour detection has many advantages comparing by event-based misbehaviour detection approaches due to its ability to early detect and stop many types of attacks in their initial stages. The Ineffectiveness issue of the existing context-based data-centric misbehaviour detection schemes is related to the inadaptability to cope with VANET harsh and high dynamic context. In this paper, a context-aware data-centric misbehaviour detection scheme was developed which contains adaptive consistency and plausibility models. These models serve as a reference that represent the vehicular context. Thus, the deviation of this context is considered misbehaviour. The experiment shows that the performance in terms of F-Measure is above 80% in all scenarios which indicate the purposed scheme is effective and robust. Advantages of the proposed solution are it works autonomously, locally in real time and privacy prevented environment. Although the CA-DC-MDS successfully detect many kinds of context-unaware attackers who manipulate the context information without considering the context situation, however, a context-aware attacker can still manipulate the context information without being detected due to the high similarity between context data and attacker's data. For example, attackers who are aware of the context boundaries can send false information but under that particular boundaries. They can incrementally adjust the context information elements. Therefore, it is important to detect such attacks in its early stage. We are currently working to solve this issue, and the finding will be reported in our future publications.

## Acknowledgement
This work was supported by the Ministry of Higher Education (MOHE) and Research Management Centre (RMC) at the Universiti Teknologi Malaysia (UTM) under Fundamental Research Grant (FRG) (VOT R.J130000.7828.4F809).

References
1. WHO. *10 facts on global road safety*. 2013 [cited 2015 12/4/2015]; Available from: http://www.who.int/features/factfiles/roadsafety/en/.
2. Arnott, R., A. de Palma, and R. Lindsey, *Economics of a bottleneck.* Journal of Urban Economics, 1990. **27**(1): p. 111-130.
3. Sweet, M., *Does traffic congestion slow the economy?* Journal of Planning Literature, 2011. **26**(4): p. 391-404.
4. Williams, B.M. and A. Guin, *Traffic Management Center Use of Incident Detection Algorithms: Findings of a Nationwide Survey.* Intelligent Transportation Systems, IEEE Transactions on, 2007. **8**(2): p. 351-358.
5. Vahdat-Nejad, H., et al., *A survey on context-aware vehicular network applications.* Vehicular Communications, 2016. **3**: p. 43-57.
6. ETSI, T., *Intelligent transport systems (ITS); vehicular communications; basic set of applications; definitions*. 2009, Tech. Rep. ETSI TR 102 638.
7. IEEE, *IEEE Guide for Wireless Access in Vehicular Environments (WAVE) - Architecture*, in *IEEE Std 1609.0-2013*. 2013. p. 1-78.
8. Lyamin, N., et al., *Real-Time Detection of Denial-of-Service Attacks in IEEE 802.11p Vehicular Networks.* Communications Letters, IEEE, 2014. **18**(1): p. 110-113.
9. Sepulcre, M., J. Gozalvez, and J. Hernandez, *Cooperative vehicle-to-vehicle active safety testing under challenging conditions.* Transportation Research Part C: Emerging Technologies, 2013. **26**: p. 233-255.
10. Uzcategui, R. and G. Acosta-Marum, *Wave: A tutorial.* Communications Magazine, IEEE, 2009. **47**(5): p. 126-133.
11. Santamaria, A.F., et al., *Road safety alerting system with radar and GPS cooperation in a VANET environment.* Wireless Sensing, Localization, and Processing Ix, 2014. **9103**.
12. Al-Sultan, S., et al., *A comprehensive survey on vehicular Ad Hoc network.* Journal of Network and Computer Applications, 2014. **37**(0): p. 380-392.
13. Hou, J., et al., *Secure and Efficient Protocol for Position-based Routing in VANETs.* 2012 Ieee International Conference on Intelligent Control, Automatic Detection and High-End Equipment (Icade), 2012: p. 142-148.
14. Williams, T., et al., *Evaluation of GPS-based methods of relative positioning for automotive safety applications.* Transportation Research Part C: Emerging Technologies, 2012. **23**: p. 98-108.
15. Milan, V., et al., *Cooperative Adaptive Cruise Control in Real Traffic Situations.* IEEE Transactions on Intelligent Transportation Systems, 2014. **15**(1): p. 296-305.
16. Heijden, R.W. and F. Kargl. *Open issues in differentiating misbehavior and anomalies for VANETs*. 2014.
17. Wymeersch, H., J. Lien, and M.Z. Win, *Cooperative Localization in Wireless Networks.* Proceedings of the IEEE, 2009. **97**(2): p. 427-450.
18. Zhang, J., *A Survey on Trust Management for VANETs*, in *25th Ieee International Conference on Advanced Information Networking and Applications (Aina 2011)*. 2011. p. 105-112.
19. Ghafoor, K., et al., *Beaconing Approaches in Vehicular Ad Hoc Networks: A Survey.* Wireless Personal Communications, 2013. **73**(3): p. 885-912.

20. Liu, K., et al., *Improving positioning accuracy using GPS pseudorange measurements for cooperative vehicular localization.* Vehicular Technology, IEEE Transactions on, 2014. **63**(6): p. 2544-2556.
21. Golestan, K., et al., *Localization in vehicular ad hoc networks using data fusion and V2V communication.* Computer Communications, 2015. **71**: p. 61-72.
22. Bissmeyer, N., W. Michael, and K. Frank, *Misbehavior detection and attacker identification in vehicular ad-hoc networks.* 2014.
23. Heijden, R.W.v.d., et al., *Survey on Misbehavior Detection in Cooperative Intelligent Transportation Systems.* arXiv preprint arXiv:1610.06810, 2016.
24. Barnwal, R.P., S.K. Ghosh, and Ieee, *Heartbeat message based misbehavior detection scheme for Vehicular Ad-hoc Networks*, in *2012 International Conference on Connected Vehicles and Expo (Iccve)*. 2012. p. 29-34.
25. Rehman, A., et al., *VANET Thread Based Message Trust Model*. 2013 Eighth International Conference on Digital Information Management, ed. I.S. Bajwa, A. Naeem, and P. Pichappan. 2013. 58-60.
26. Heijden, R.W.v.d., et al., *Survey on Misbehavior Detection in Cooperative Intelligent Transportation Systems.* CoRR, 2016. **abs/1610.06810**.
27. Leinmuller, T., et al., *Decentralized position verification in geographic ad hoc routing.* Security and Communication Networks, 2010. **3**(4): p. 289-302.
28. van der Heijden, R., S. Dietzel, and F. Kargl, *Misbehavior detection in vehicular ad-hoc networks*. 2013.
29. Khan, U., S. Agrawal, and S. Silakari, *A detailed survey on misbehavior node detection techniques in vehicular ad Hoc networks*, in *Advances in Intelligent Systems and Computing*. 2015. p. 11-19.
30. van der Heijden, R.W., et al., *Survey on Misbehavior Detection in Cooperative Intelligent Transportation Systems.* arXiv preprint arXiv:1610.06810, 2016.
31. Van der Heijden, R.W., et al. *Enhanced position verification for VANETs using subjective logic*. in *Proceedings of the 2016 IEEE 84th Vehicular Technology Conference*. 2016. Montreal, Canada: Universität Ulm.
32. Sakiz, F. and S. Sen, *A survey of attacks and detection mechanisms on intelligent transportation systems: VANETs and IoV.* Ad Hoc Networks, 2017. **61**: p. 33-50.
33. Golle, P., D. Greene, and J. Staddon, *Detecting and correcting malicious data in VANETs*, in *Proceedings of the 1st ACM international workshop on Vehicular ad hoc networks*. 2004, ACM: Philadelphia, PA, USA. p. 29-37.
34. Raya, M., et al. *On Data-Centric Trust Establishment in Ephemeral Ad Hoc Networks*. in *INFOCOM 2008. The 27th Conference on Computer Communications. IEEE*. 2008.
35. Chen, Y.-M. and Y.-C. Wei, *A beacon-based trust management system for enhancing user centric location privacy in VANETs.* Communications and Networks, Journal of, 2013. **15**(2): p. 153-163.
36. Leinmuller, T. and E. Schoch. *Greedy routing in highway scenarios: The impact of position faking nodes*. in *Proceedings of Workshop On Intelligent Transportation (WIT 2006)(Mar. 2006)*. 2006.
37. Jaeger, A., et al., *A Novel Framework for Efficient Mobility Data Verification in Vehicular Ad-hoc Networks.* International Journal of Intelligent Transportation Systems Research, 2012. **10**(1): p. 11-21.
38. Zaidi, K., et al., *Host-Based Intrusion Detection for VANETs: A Statistical Approach to Rogue Node Detection.* IEEE Transactions on Vehicular Technology, 2016. **65**(8): p. 6703-6714.
39. Dietzel, S., et al., *A flexible, subjective logic-based framework for misbehavior detection in V2V networks*, in *World of Wireless, Mobile and Multimedia Networks (WoWMoM), 2014 IEEE 15th International Symposium on a*. 2014. p. 1-6.
40. Nai-Wei, L. and T. Hsiao-Chien. *Illusion Attack on VANET Applications - A Message Plausibility Problem*. in *Globecom Workshops, 2007 IEEE*. 2007.
41. Grover, J., M.S. Gaur, and V. Laxmi. *Position Forging Attacks in Vehicular Ad Hoc Networks: Implementation, Impact and Detection*. in *Wireless Communications and Mobile Computing Conference (IWCMC), 2011 7th International*. 2011.
42. Hsiao, H.-C., et al., *Flooding-resilient broadcast authentication for VANETs*, in *Proceedings of the 17th annual international conference on Mobile computing and networking*. 2011, ACM: Las Vegas, Nevada, USA. p. 193-204.
43. Palomar, E., et al., *Hindering false event dissemination in VANETs with proof-of-work mechanisms.* Transportation Research Part C-Emerging Technologies, 2012. **23**: p. 85-97.
44. Yu, B., C.-Z. Xu, and B. Xiao, *Detecting Sybil attacks in VANETs.* Journal of Parallel and Distributed Computing, 2013. **73**(6): p. 746-756.
45. Bissmeyer, N., et al., *Assessment of node trustworthiness in vanets using data plausibility checks with particle filters*, in *Vehicular Networking Conference (VNC)*. 2012, 2012 IEEE: Seoul, 2012. p. pp. 78-85.
46. Firl, J., et al., *MARV-X: Applying Maneuver Assessment for Reliable Verification of Car-to-X Mobility Data.* Intelligent Transportation Systems, IEEE Transactions on, 2013. **14**(3): p. 1301-1312.
47. Hubaux, J.-P., S. Capkun, and J. Luo, *The security and privacy of smart vehicles.* IEEE Security & Privacy Magazine, 2004. **2**(LCA-ARTICLE-2004-007): p. 49--55.
48. Leinmüller, T., et al. *Improved security in geographic ad hoc routing through autonomous position verification*. in *VANET - Proceedings of the Third ACM International Workshop on Vehicular Ad Hoc Networks*. 2006.
49. Demirbas, M. and Y. Song, *An RSSI-based Scheme for Sybil Attack Detection in Wireless Sensor Networks*, in *Proceedings of the 2006 International Symposium on on World of Wireless, Mobile and Multimedia Networks*. 2006, IEEE Computer Society. p. 564-570.
50. Schmidt, R.K., et al. *Vehicle behavior analysis to enhance security in vanets*. in *Proceedings of the 4th IEEE Vehicle-to-Vehicle Communications Workshop (V2VCOM2008)*. 2008.
51. Bissmeyer, N., C. Stresing, and K.M. Bayarou. *Intrusion detection in VANETs through verification of vehicle movement data*. in *Vehicular Networking Conference (VNC), 2010 IEEE*. 2010.
52. Parker, R. and S. Valaee, *Vehicular Node Localization Using Received-Signal-Strength Indicator.* Vehicular Technology, IEEE Transactions on, 2007. **56**(6): p. 3371-3380.
53. Stübing, H., et al. *Verifying mobility data under privacy considerations in car-to-x communication*. in *17th ITS World Congress*. 2010.

54. Ghaleb, F.A., et al., *Improved vehicle positioning algorithm using enhanced innovation-based adaptive Kalman filter.* Pervasive and Mobile Computing, 2017. **40**: p. 139-155.
55. Grover, J., et al., *Machine Learning Approach for Multiple Misbehavior Detection in VANET.* Advances in Computing and Communications, Pt Iii, 2011. **192**: p. 644-653.
56. Yan, G.J., S. Olariu, and M.C. Weigle, *Providing VANET security through active position detection.* Computer Communications, 2008. **31**(12): p. 2883-2897.
57. Bissmeyer, N., et al., *Central misbehavior evaluation for VANETs based on mobility data plausibility*, in *Proceedings of the ninth ACM international workshop on Vehicular inter-networking, systems, and applications*. 2012, ACM: Low Wood Bay, Lake District, UK. p. 73-82.
58. Malandrino, F., et al., *Verification and Inference of Positions in Vehicular Networks through Anonymous Beaconing.* Mobile Computing, IEEE Transactions on, 2014. **13**(10): p. 2415-2428.
59. Leinmueller, T., et al., *Decentralized position verification in geographic ad hoc routing.* Security and Communication Networks, 2010. **3**(4): p. 289-302.
60. Karagiannis, G., et al., *Vehicular Networking: A Survey and Tutorial on Requirements, Architectures, Challenges, Standards and Solutions.* Communications Surveys & Tutorials, IEEE, 2011. **13**(4): p. 584-616.
61. Raya, M., et al., *Eviction of Misbehaving and Faulty Nodes in Vehicular Networks.* Selected Areas in Communications, IEEE Journal on, 2007. **25**(8): p. 1557-1568.
62. Wahab, O.A., H. Otrok, and A. Mourad, *A cooperative watchdog model based on Dempster-Shafer for detecting misbehaving vehicles.* Computer Communications, 2014. **41**: p. 43-54.
63. Tian, X.Y., et al., *Computational Security for Context-Awareness in Vehicular Ad-Hoc Networks.* IEEE Access, 2016. **4**: p. 5268-5279.
64. Wahab, O.A., et al., *CEAP: SVM-based intelligent detection model for clustered vehicular ad hoc networks.* Expert Systems with Applications, 2016. **50**(Supplement C): p. 40-54.
65. FHWA. *Next Generation Simulation (NGSIM) Vehicle Trajctories Dataset*. 2006 17/12/2015; 2006:[Available from: http://ngsim-community.org/.
66. Hou, Y., P. Edara, and C. Sun, *Modeling Mandatory Lane Changing Using Bayes Classifier and Decision Trees.* IEEE Transactions on Intelligent Transportation Systems, 2014. **15**(2): p. 647-655.
67. Hartigan, J.A. and M.A. Wong, *Algorithm AS 136: A k-means clustering algorithm.* Journal of the Royal Statistical Society. Series C (Applied Statistics), 1979. **28**(1): p. 100-108.
68. Drawil, N.M. and O. Basir, *Intervehicle-Communication-Assisted Localization.* IEEE Transactions on Intelligent Transportation Systems, 2010. **11**(3): p. 678-691.
69. Ghaleb, A.F., et al., *Driving-situation-aware adaptive broadcasting rate scheme for vehicular ad hoc network.* Journal of Intelligent & Fuzzy Systems. **35**(1): p. 1-16.
70. Bissmeyer, N., et al., *Short paper: Experimental analysis of misbehavior detection and prevention in VANETs*, in *Fifth IEEE Vehicular Networking Conference, VNC 2013*. 2013, IEEE Communications Society: Boston. p. 198-201.
71. Nikaein, N., et al., *Application Distribution Model and Related Security Attacks in VANET.* International Conference on Graphic and Image Processing (Icgip 2012), 2013. **8768**.